\documentclass[sigconf,nonacm]{acmart}

\renewcommand\footnotetextcopyrightpermission[1]{}
\AtBeginDocument{}

\usepackage{algorithm}
\usepackage{graphicx}
\usepackage{algpseudocode}
\usepackage{enumitem}
\newcommand{\E}{\mathbb{E}}
\newcommand{\KL}{\mathrm{KL}}

\newcommand{\EpochsSeq}{2}
\newcommand{\EpochsKR}{2}
\newcommand{\UpdatesSynth}{2}
\providecommand{\TBD}{\textbf{TBD}}

\begin{document}

\title{Learning Generative Ranking Policies for Whole-Page Optimization}

\author{\parbox{\textwidth}{\centering
Chen Qiu\textsuperscript{\textdagger} \quad Zhigang Hua\textsuperscript{\textdagger} \quad Qi Xu \quad Jizhe Zhang \quad Sha Hua \quad Ren Mao \quad Luke Simon\textsuperscript{\textdagger}\\[6pt]}}
\affiliation{\institution{Meta Platforms, Inc.}\country{United States}}

\begin{abstract}
A recommendation page is consumed as a whole: the items on it compete for attention and complement one another, so the best page is not simply the individually best items placed side by side. We study whole-page optimization, which selects and orders the items on a page jointly to maximize the score of the page. The number of possible pages grows combinatorially, so good pages must be generated rather than enumerated. Some generative re-ranking methods use the same page-level training signal for every choice in a sampled page. Others imitate target pages built before training, which remain largely fixed as the generator improves. We address both limitations with one local page-level objective, which repeatedly moves the current generator toward pages with higher score. This objective splits exactly into one training target per position, and each target compares candidates at the same partial page by the scores of the complete pages they lead to. Every position therefore learns to choose the item that is better for the whole page, rather than the item that looks best on its own. We call this procedure whole-page policy improvement. On two public datasets, KuaiRand-1K and RL4RS-SeqSlate, which cover multi-objective item selection and constrained ordering, and on an industrial dataset of logged pages, our method outperforms the evaluated generative re-ranking baselines. On a synthetic benchmark with a known optimal page, it also comes closest to the optimum, showing that its improvements come from better whole-page decisions.

\end{abstract}



\maketitle
\begingroup
\renewcommand{\thefootnote}{\textdagger}
\footnotetext{Corresponding author: Chen Qiu (chenqiu@meta.com), Zhigang Hua (zhua@meta.com), Luke Simon (lukesimon@meta.com)}
\endgroup
\pagestyle{plain}

\section{Introduction}\label{sec:intro}
Web platforms such as content feeds, e-commerce sites and video services serve recommendations as pages of items that users browse together.
The contribution of a recommendation item depends on the page around it.
An early selection affects both the candidates that remain feasible and the context of later recommendations.
A recommendation system must therefore select and order items jointly, while balancing engagement objectives and satisfying page constraints.
Whole-page optimization seeks to maximize page score through this joint selection and ordering decision~\cite{wang2016beyondranking,zhao2018deeppage}.

The space of feasible pages grows combinatorially, so pages cannot be enumerated, and evaluator search at serving time requires many queries for every request.
A common approach is to train a generator against a page evaluator and use the learned generator for serving~\cite{goalrank,degre}.
We take the evaluator as given and focus on generator training.
Generative re-ranking methods either learn from scores of sampled pages, which give one page-level signal to every choice on the page, or fit target lists built by an evaluator, which remain largely fixed as the generator improves~\cite{liu2023gfn4rec,goalrank,degre,wang2025nlgr}.
The training challenge is to compare next-item alternatives at the same partial page by the complete pages they lead to, and to keep these comparisons current as the generator changes.

We propose \emph{whole-page policy improvement}, a training procedure that alternates search guided by the current generator with learning from complete-page scores.
We turn those scores into per-position targets, so each choice is trained according to the whole page it leads to.
The updated generator guides the next round, allowing both the search and the targets to adapt as training progresses.
We derive this procedure from a local page-score objective and prove improvement for exact updates and convergence to the optimal page score for repeated exact full updates.
At serving time, the trained generator produces pages without evaluator search.

\begin{figure*}[t]
\centering
\includegraphics[width=\textwidth]{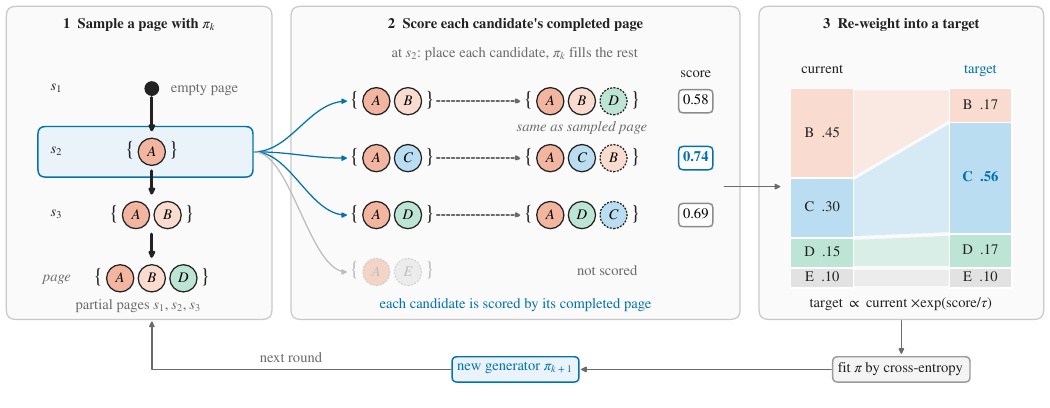}
\caption{One round of whole-page policy improvement on a toy request.
At the partial page containing A, each scored candidate is placed next and $\pi_k$ samples the remaining item (dotted circles).
The resulting complete-page scores redistribute probability within the scored set, preserving its total mass and all unscored probabilities.
Cross-entropy fitting produces $\pi_{k+1}$, which constructs the next round's data and targets.}
\label{fig:overview}
\end{figure*}

Our method outperforms generative recommendation baselines and representative RL algorithms at every length setting of two public datasets and on an industrial dataset of logged pages.
Compared with prefix-score beam search, the trained generator achieves better page quality at every tested width and setting, with much lower measured average generation time and no evaluator queries.
On a synthetic benchmark with known page scores, it also comes closest to the optimum without a learned referee.
Ablations on the public datasets show that the improvements come from scoring candidates through complete pages and from constructing training data with the updated generator.
Our contributions are:
\begin{itemize}[leftmargin=*]
  \item We introduce whole-page policy improvement, which evaluates next-item alternatives through pages completed by the current generator and turns the resulting scores into per-position supervision for generative re-ranking.
  \item We derive closed-form per-position targets from a local page-score objective, including a partial-scoring update that redistributes probability among scored candidates, and prove that exact updates improve expected page score and that repeated exact full updates converge to the optimal page score.
  \item We demonstrate consistent improvements on two public datasets and an industrial dataset, with better page quality at much lower measured generation time than prefix-score beam search. On a synthetic benchmark, our method also comes closest to the optimum under known page scores, and ablations show that the improvements come from complete-page candidate scoring and from constructing training data with the updated generator.
\end{itemize}

\section{Related Work}\label{sec:related}

\paragraph{Listwise and whole-page recommendation.}
Beyond Ranking~\cite{wang2016beyondranking} studies whole-page presentation optimization.
Listwise recommendation models account for dependencies among items shown together.
PRM~\cite{pei2019prm} models personalized list context, while Seq2Slate~\cite{seq2slate} conditions each choice on the items already selected.
LIGE-GR~\cite{srinivas2026ligegr} uses a context-aware predictor and a listwise value model for greedy or beam decoding at serving time.
DeepPage~\cite{zhao2018deeppage} jointly generates complementary items and their two-dimensional display.
PageLLM~\cite{pagellm} combines page-level and item-level rewards to train a page-generating language model.
SlateQ~\cite{ie2019slateq} decomposes slate values under user-choice assumptions; we derive next-item targets from pages completed by the current generator.

\paragraph{Generative recommendation.}
Generative recommenders produce item sequences or slates in several ways.
TIGER~\cite{rajput2023tiger} predicts the next item's semantic identifier, while HSTU~\cite{zhai2024hstu} formulates recommendation as sequential transduction.
List-CVAE~\cite{listcvae} models a joint slate distribution conditioned on user responses.
Prompt-to-Slate~\cite{tomasi2025prompttoslate} uses diffusion to generate slates conditioned on text prompts.
GFN4Rec~\cite{liu2023gfn4rec} aligns list-generation probabilities with rewards, and HAC~\cite{liu2023hac} generates lists through latent hyper-actions.
OneRec~\cite{deng2025onerec} combines session-wise list generation with iterative DPO alignment.
HiGR~\cite{pang2025higr} combines hierarchical slate decoding with ORPO-based listwise alignment.
OneRec and HiGR use list-level preference alignment, while RGD~\cite{yang2026rgd} reweights token probabilities with a learned reward during decoding; we construct per-position training targets from complete-page comparisons.

\paragraph{Generative re-ranking.}
Earlier evaluator--generator frameworks train list generators against contextual value estimates~\cite{huzhang2020egrerank,wei2020generatorcritic,grn}.
PIER~\cite{shi2023pier} learns permutation selection and contextual evaluation jointly.
NAR4Rec~\cite{nar4rec} uses non-autoregressive generation, while DCDR~\cite{lin2023dcdr} uses discrete diffusion.
Multi-generator re-ranking~\cite{multi_generator_reranking} learns complementary list proposals.
SORT-Gen~\cite{sortgen} combines sub-list value estimation with efficient generation for multiple objectives.
GReF~\cite{lin2025gref} combines Rerank-DPO with ordered multi-token prediction.
NLGR~\cite{wang2025nlgr} fits one-item neighbor edits by reward-weighted likelihood with the other items fixed; we re-complete the remaining page with the current generator.
DeGRe~\cite{degre} distills fixed targets constructed by offline beam search and GoalRank~\cite{goalrank} trains with groups of whole lists; we rebuild per-position targets from pages completed by the updated generator.

\section{Methodology}\label{sec:method}

We first formulate whole-page optimization and outline the training procedure.
We then derive the per-position targets, describe the learning algorithm, and analyze the improvement and convergence.

\subsection{Problem formulation}\label{sec:formulation}

Given a request $c$ with user context and candidate pool $\mathcal C$, we generate an ordered page $x=(a_1,\ldots,a_L)$ by selecting $L$ candidates without replacement.
The page must satisfy the task's selection and placement constraints.

The generator builds the page one item at a time.
At position $t$, the generator observes the partial page $s_t=(c,a_1,\ldots,a_{t-1})$.
The legal set $\mathcal A(s_t)$ contains unused candidates that permit a feasible full-page completion.
The generator defines a distribution $\pi(\cdot\mid s_t)$ over these legal candidates.
The probability of a complete page is therefore $\pi(x\mid c)=\prod_{t=1}^{L}\pi(a_t\mid s_t)$.

A page evaluator assigns a score $r(x)$ to each complete page for its request.
We call $r(x)$ the page score; expectations of $r$ are called expected scores.
We omit $c$ from the evaluator notation.
We take the evaluator as given and keep it fixed during generator training.
For recommendation logs, the evaluator is learned before generator training.
Our objective is to maximize the expected page score, $J(\pi)=\E_c\,\E_{x\sim\pi(\cdot\mid c)}[r(x)]$.

\subsection{Whole-page policy improvement}\label{sec:wppi}

Whole-page policy improvement trains each next-item choice by comparing the complete pages that alternative candidates lead to (Figure~\ref{fig:overview}).
Training proceeds in rounds.
In round $k$, we hold the current generator $\pi_k$ fixed while constructing targets.

First, $\pi_k$ samples pages to obtain partial pages.
At each partial page, we place each scored candidate next, use $\pi_k$ to sample the remaining items, and score the resulting complete page.
These scores reweight $\pi_k$'s next-item probabilities into a position-specific target.
Each scored candidate is evaluated through the complete page that follows its selection.
The score accounts for the candidate's placement and the resulting continuation.
In Figure~\ref{fig:overview}, $\pi_k$ favors B after A, but the completed page ACB through C scores higher than the sampled page ABD.
The target shifts probability from B to C.
We fit the generator to these targets by cross-entropy, then use the fitted generator to construct the next round's data and targets.

\subsection{Target construction}\label{sec:targets}

We derive a training target for each position from the expected score of the complete page.

\paragraph{Candidate scores and page improvement.}
To compare candidates at the same partial page $s$, we hold the prefix fixed, place candidate $a$ next, and let $\pi_k$ generate the remaining items.
Write $\pi_k(\cdot\mid s,a)$ for the resulting distribution of complete pages.
The candidate score is the expected evaluator score of these pages:
$Q^{\pi_k}(s,a)=\E_{x\sim\pi_k(\cdot\mid s,a)}[r(x)]$.
The evaluator scores the entire completed page, including the fixed prefix, the candidate, and the sampled continuation.
Thus, each candidate is scored by the whole page produced after choosing it.

If $\pi_k$ also chooses the next item, its expected page score from $s$ is
$V^{\pi_k}(s)=\sum_a\pi_k(a\mid s)Q^{\pi_k}(s,a)$.
We use $V^{\pi_k}(s)$ as the baseline for comparing candidate choices.
Choosing $a$ next changes this expected score by $Q^{\pi_k}(s,a)-V^{\pi_k}(s)$, the advantage of $a$, with the remaining items still generated by $\pi_k$.

Let $\pi$ be a possible update to $\pi_k$.
At $s$, we let $\pi$ choose the next item and keep $\pi_k$ for the remaining items.
The expected advantage of $\pi$ at $s$ is
$\sum_a\pi(a\mid s)Q^{\pi_k}(s,a)-V^{\pi_k}(s)$.
Updating the generator also changes the probabilities of reaching different partial pages.
Let $d_t^\pi$ denote the distribution of partial pages before selecting the item at position $t$ under $\pi$.
The performance difference identity~\cite{kakade2002approximately} says that the expected advantages at each position sum to the total change in $J$:
\begin{equation}\label{eq:pdl}
\begin{aligned}
J(\pi)-J(\pi_k)
={}&\sum_{t=1}^{L}\E_{s\sim d_t^\pi}
\Big[\sum_a\pi(a\mid s)Q^{\pi_k}(s,a)
-V^{\pi_k}(s)\Big].
\end{aligned}
\end{equation}
Appendix~\ref{app:proofs} gives the finite-horizon proof (Lemma~\ref{lem:pdl}).

\paragraph{A local training objective.}
In~\eqref{eq:pdl}, the partial-page distribution $d_t^\pi$ changes with the proposed generator $\pi$.
Our training data, however, are sampled from the reference generator $\pi_k$.
We replace $d_t^\pi$ with $d_t^{\pi_k}$ so the partial-page weights remain fixed during target construction.
We also add a KL penalty to discourage large changes from $\pi_k$.
Write $\KL_s(\pi\|\pi_k)=\KL(\pi(\cdot\mid s)\|\pi_k(\cdot\mid s))$.
The local objective is
\begin{equation}\label{eq:surrogate}
\begin{aligned}
\mathcal L_k(\pi)
={}&J(\pi_k)+\sum_{t=1}^{L}\E_{s\sim d_t^{\pi_k}}
\Big[\sum_a\pi(a\mid s)Q^{\pi_k}(s,a)\\
&\hspace{3em}-V^{\pi_k}(s)-\tau\,\KL_s(\pi\|\pi_k)\Big].
\end{aligned}
\end{equation}
At $\pi=\pi_k$, $\mathcal L_k$ and $J$ coincide, as do their gradients (Appendix~\ref{app:surrogate}).
A small change from $\pi_k$ therefore has the same first-order effect under both objectives.
The target temperature $\tau>0$ controls how strongly the KL term penalizes changes from $\pi_k$.

\paragraph{Per-position targets.}
With the partial-page weights and candidate scores fixed, $\mathcal L_k$ separates into one optimization for each next-item distribution.
The terms $J(\pi_k)$ and $V^{\pi_k}(s)$ are constant, and each conditional distribution $\pi(\cdot\mid s)$ enters only the term for $s$, weighted by its visitation probability $d_t^{\pi_k}(s)$.
Over free conditional distributions, we can therefore solve each term separately.
The target is position-specific, but its candidate scores still account for the complete page.
At a partial page $s$, the problem is
\begin{equation}\label{eq:perstate}
\max_{\pi(\cdot\mid s)}\;
\sum_{a\in\mathcal A(s)}\pi(a\mid s)Q^{\pi_k}(s,a)
-\tau\,\KL_s(\pi\|\pi_k).
\end{equation}
This problem has a closed-form solution, also used in MPO~\cite{abdolmaleki2018maximum}.
The solution re-weights the current probabilities by $\exp(Q^{\pi_k}(s,a)/\tau)$, then normalizes the weights across legal candidates.
We denote the resulting target distribution by $q_k(\cdot\mid s)$.

\begin{proposition}[Per-position target]\label{prop:split}
The unique solution of~\eqref{eq:perstate} is
\begin{equation}\label{eq:target}
q_k(a\mid s)=
\frac{\pi_k(a\mid s)\exp\!\left(Q^{\pi_k}(s,a)/\tau\right)}
{\sum_{b\in\mathcal A(s)}\pi_k(b\mid s)\exp\!\left(Q^{\pi_k}(s,b)/\tau\right)}.
\end{equation}
Choosing $q_k(\cdot\mid s)$ at every partial page maximizes $\mathcal L_k$.
\end{proposition}
The proof is in Appendix~\ref{app:split}.

Candidates with higher $Q^{\pi_k}(s,a)$ receive larger multiplicative weights.
Candidates with equal $Q^{\pi_k}(s,a)$ retain their current probability ratio.
These comparisons account for the complete page, with the remaining items generated by $\pi_k$.

\paragraph{Round after round.}
The targets are defined with respect to the current generator $\pi_k$.
After the generator is updated, the next round re-evaluates it, recomputing candidate scores from its continuations, and builds targets at the partial pages it now visits, so the targets keep pace with the generator.
Section~\ref{sec:theory} proves that exact target updates improve expected page score and that repeated exact full updates converge to the optimal page score.

\subsection{Learning algorithm}\label{sec:algorithm}

We now turn the target construction into a training procedure (Algorithm~\ref{alg:wppi}).
Each round forms Monte Carlo estimates of $Q^{\pi_k}$ from sampled completions, constructs target distributions, and fits the generator by cross-entropy.

\paragraph{Sampling and scoring.}
Each round selects a batch of training requests.
For each selected request, $\pi_k$ samples one page of length $L$.
We collect the partial page before each of the $L$ item choices.
These are $s_1,\ldots,s_L$; $s_1$ contains only the request, before any item is placed.
Scoring every legal candidate can be resource intensive.
At each partial page $s$, we choose a scored set $S(s)$ before querying the evaluator.
The set contains the top $\kappa$ candidates by $\pi_k$, plus $\kappa_{\mathrm r}$ candidates drawn uniformly from the remaining legal candidates.
We score all legal candidates when their number is at most $\kappa+\kappa_{\mathrm r}$.

For each scored candidate $a$, we place $a$ next and let $\pi_k$ sample one completion $x'$.
We set $\hat Q(s,a)=r(x')$.
In the rollout model, $\E[\hat Q(s,a)]=Q^{\pi_k}(s,a)$.
These rollout scores are unbiased Monte Carlo estimates of $Q^{\pi_k}$; no critic is learned.

\paragraph{Constructing targets.}
We apply the score-based reweighting within the scored set $S(s)$, while keeping every unscored candidate's probability unchanged.
This fixes the total scored probability mass at $m(s)=\sum_{a\in S(s)}\pi_k(a\mid s)$.
We redistribute that mass among scored candidates using their estimated complete-page scores:
\begin{equation}\label{eq:renorm}
\hat q_k(a\mid s)=
\begin{cases}
\pi_k(a\mid s), & a\notin S(s),\\[3pt]
\dfrac{m(s)\pi_k(a\mid s)e^{\hat Q(s,a)/\tau}}
{\sum_{b\in S(s)}\pi_k(b\mid s)e^{\hat Q(s,b)/\tau}},
& a\in S(s).
\end{cases}
\end{equation}
If $m(s)=0$, we define $\hat q_k(\cdot\mid s)=\pi_k(\cdot\mid s)$.
When all legal candidates are scored, \eqref{eq:renorm} is~\eqref{eq:target} with $\hat Q$ in place of $Q^{\pi_k}$.

\paragraph{Fitting the generator.}
We fit a parametric generator $\pi_\theta$ to the targets using the sampled partial pages.
The per-position cross-entropy objective is
\begin{equation}\label{eq:ce}
\ell_k(\theta)=
-\sum_{t=1}^{L}\E_{s\sim d_t^{\pi_k}}
\sum_a\hat q_k(a\mid s)\log\pi_\theta(a\mid s).
\end{equation}
Cross-entropy fits the parametric generator by minimizing $\KL(\hat q_k\|\pi_\theta)$.
The exact local objective instead minimizes $\KL(\pi\|q_k)$.
With full exact targets and free conditional distributions, both are minimized by matching $q_k$.
For the parametric generator, we use the cross-entropy fit as the projection onto the target distributions.
We initialize the fit with $\pi_k$ and make a small number of passes over the fixed targets.
The fitted generator becomes $\pi_{k+1}$ and starts the next round.
At serving time, it selects the highest-probability legal candidate at each position, without evaluator queries or sampled completions.

\begin{algorithm}[t]
\caption{Whole-page policy improvement}\label{alg:wppi}
\small
\begin{algorithmic}[1]
\setlength{\itemsep}{0pt}
\setlength{\parsep}{0pt}
\Statex \textbf{Require:} initial generator $\pi_0$; evaluator $r$;
\Statex \phantom{\textbf{Require:}\ }training requests $\mathcal D$; temperature $\tau$
\For{$k=0,1,2,\ldots$}
  \State select a batch of requests from $\mathcal D$
  \State sample one page per selected request from $\pi_k$
  \State collect partial pages from the sampled pages
  \ForAll{collected partial pages $s$}
    \State choose the scored set $S(s)$
    \State $\hat Q(s,a)\gets r(x')$, $x'\sim\pi_k(\cdot\mid s,a)$, for $a\in S(s)$
    \State construct $\hat q_k(\cdot\mid s)$ using~\eqref{eq:renorm}
  \EndFor
  \State fit $\pi_{k+1}$ by~\eqref{eq:ce}, starting from $\pi_k$
\EndFor
\end{algorithmic}
\end{algorithm}

\subsection{Properties of the update}\label{sec:theory}

An exact update uses $Q^{\pi_k}$ and sets the generator to the target distributions at every partial page.
A full update scores every legal candidate; a partial update keeps unscored probabilities fixed.
Algorithm~\ref{alg:wppi} uses sampled $\hat Q$ and a finite cross-entropy fit to approximate these updates.

\paragraph{Exact target updates improve page score.}
Applying the exact targets at every partial page defines an updated generator $q_k$.
The next result shows this update never decreases expected page score.

\begin{proposition}[Improvement]\label{prop:improve}
For any $\tau>0$,
\begin{equation}\label{eq:improve}
\begin{aligned}
J(q_k)-J(\pi_k)
={}&\tau\sum_{t=1}^{L}\E_{s\sim d_t^{q_k}}
\Big[\KL_s(q_k\|\pi_k)\\
&\hspace{5em}+\KL_s(\pi_k\|q_k)\Big]\ge0.
\end{aligned}
\end{equation}
\end{proposition}
The proof is in Appendix~\ref{app:improve}.
At each partial page, $q_k$ improves the expected page score obtained by drawing the next item from $q_k$ and the remaining items from $\pi_k$, since keeping $\pi_k$ is feasible in~\eqref{eq:perstate}.
Since this holds at every partial page, the sum in~\eqref{eq:pdl} remains nonnegative under the updated generator's partial-page distribution, which closes the gap left by fixing $d_t^{\pi_k}$ in $\mathcal L_k$.
The improvement equals $\tau$ times a symmetric KL divergence, so it is positive whenever the target differs from $\pi_k$.
The guarantee holds at every positive temperature and requires no small-step assumption.

\paragraph{Partial scoring preserves improvement.}
Keeping the unscored probabilities fixed makes partial scoring a constrained version of~\eqref{eq:perstate}.
Let $q_k^S$ denote~\eqref{eq:renorm} with $Q^{\pi_k}$ in place of $\hat Q$.

\begin{proposition}[Partial scoring]\label{prop:renorm}
For any fixed scored sets $S(s)$, $q_k^S(\cdot\mid s)$ maximizes~\eqref{eq:perstate} subject to keeping every unscored probability fixed as under $\pi_k$.
Moreover,
\begin{equation}\label{eq:renorm-improve}
J(q_k^S)-J(\pi_k)\ge0.
\end{equation}
\end{proposition}
The proof is in Appendix~\ref{app:renorm}.
With exact $Q^{\pi_k}$, reweighting only the scored set retains policy improvement because the unscored probabilities stay fixed and the scored set's expected score does not decrease.
The guarantee holds for any choice of scored sets; scoring fewer candidates can only make each round's step smaller, not reverse its direction.

\paragraph{Repeated exact updates converge to the optimal page score.}
The previous results establish improvement within a round.
Repeating the full target update also connects these local decisions to the best complete page.
The exact iteration is an instance of policy mirror descent~\cite{geist2019theory,shani2020adaptive}.

\begin{proposition}[Convergence]\label{prop:convergence}
Assume that each request has finitely many feasible pages and that evaluator scores are bounded.
Let $\pi_{k+1}=q_k$ at every round, with a fixed $\tau>0$.
For each request, the expected page score converges to the best page score within the initial generator's support.
If $\pi_0$ assigns positive probability to every legal choice, the limit is the global optimum under the evaluator.
\end{proposition}
The proof is in Appendix~\ref{app:convergence}.
Repeated exact full updates thus improve expected page score toward the best supported page.
Our $\pi_0$ is a softmax over legal candidates and supports every legal choice, so this limit is the global optimum under the evaluator; the gap to any supported page shrinks at rate $O(1/K)$.

\section{Experiments}\label{sec:experiments}
\providecommand{\TBD}{\textbf{TBD}}

We evaluate whole-page policy improvement in four parts.
On two public datasets with six length settings, we compare it with generator-training baselines (\S\ref{sec:main-results}) and with prefix-score beam search at serving time (\S\ref{sec:serving-search}); an industrial dataset of logged pages tests the same two comparisons.
Because scores on the public and industrial datasets come from a learned referee, a synthetic benchmark with known page scores and exact optima evaluates the methods without a learned referee, and allows a decomposition of page score that these datasets cannot provide (\S\ref{sec:exp-synth}).
Finally, ablations on the public datasets identify the components behind the improvements (\S\ref{sec:ablations}).

\subsection{Setup}\label{sec:setup}

\paragraph{Datasets.}
\begingroup
\setlength{\parindent}{0pt}
\setlength{\parskip}{3pt}
We use two public datasets, an industrial dataset and a synthetic benchmark with exact optimal pages.
Table~\ref{tab:dataset-space} lists the data used and the feasible-page space at each length setting.

\textbf{KuaiRand-1K~\cite{gao2022kuairand}.}
We use its feed logs for multi-objective item selection and ordering from expanded candidate pools.
A page joins 2, 4 or 8 consecutive requests from one user within a session, with sessions separated by gaps of 30 minutes.
The reported $L\in\{15,28,54\}$ denotes the model position capacity; pages contain about 9, 18 and 36 real items on average.
For a page containing $n$ logged items, the candidate pool includes those items and up to $n$ additional items shown to the user in the next 24 hours.
The pools contain about 17, 35 and 70 candidates on average; the generator fills $n\le L$ real positions, with padding masked.
Using items from future exposures makes this a retrospective offline benchmark, with the same candidate pools for every method.
Page score sums predicted clicks, long views and likes, with weights 1, 1 and 10.
Training and test use the full date splits.

\textbf{RL4RS-SeqSlate~\cite{wang2021rl4rs}.}
We use its purchase logs for constrained reordering of fixed candidate pools.
A page joins 2, 3 or 4 of a session's 9-item slates, giving $L\in\{18,27,36\}$.
The generator reorders all $L$ logged pool entries.
Each entry can occupy its assigned row type (top, middle or bottom) in any slate; item IDs cannot repeat within a slate, and each slate holds at most one special item (Appendix~\ref{app:dataset-constraints}).
Page score sums predicted purchase probability times an item-level weight.
Training and test use the full valid folds.

\textbf{Industrial dataset.}
We use logs from a large-scale content feed.
A page is one served request of up to $L=20$ items, mixing organic and sponsored items.
The generator reorders the page's own served items without adding candidates.
Sponsored items must satisfy the page's spacing rule: a minimum number of organic items before the first sponsored item and between consecutive sponsored items, except for a limited number of adjacent pairs.
Page score sums predicted outcomes from sponsored items and predicted engagement from organic items.
Training and validation use 199{,}885 pages sampled from one day (5{,}000 for validation); test uses pages sampled from each of the following three days (about 80{,}000 per day), and results average the three test days.

\textbf{Synthetic.}
Synthetic data are regenerated for each seed, and all methods share the resulting item catalog and requests.
Every method orders the same 10 candidates for each request.
There are $10!\approx3.6$M feasible pages, so enumeration gives the exact optimum.
Ground-truth page score is the expected total utility of clicked items.
Click probabilities depend on position and preceding items through format-by-position fit, topic redundancy and complementarity, and same-format fatigue.
Browsing can stop after a satisfying click or a screen without clicks.
Methods train against this ground-truth evaluator on 6{,}000 training requests and are tested on 3{,}000 (Appendix~\ref{app:synthetic-details}).
\par
\endgroup

\begin{table}[t]
\centering
\caption{Dataset sizes and feasible-page counts (median distinct item-ID sequences per request).
Length is the position capacity for KuaiRand-1K and the industrial dataset; other datasets use real page length.
}\label{tab:dataset-space}
\small
\setlength{\tabcolsep}{3pt}
\begin{tabular*}{\columnwidth}{@{\extracolsep{\fill}}crrc@{}}
\toprule
Length & Train & Test & Feasible pages \\
\midrule
\multicolumn{4}{l}{KuaiRand-1K} \\
$\leq 15$ & $606{,}723$ & $110{,}235$ & $\sim 10^{10}$ \\
$\leq 28$ & $278{,}651$ & $50{,}766$ & $\sim 10^{26}$ \\
$\leq 54$ & $119{,}888$ & $21{,}932$ & $\sim 10^{62}$ \\
\midrule
\multicolumn{4}{l}{RL4RS-SeqSlate} \\
$18$ & $279{,}396$ & $224{,}340$ & $\sim 10^{8}$ \\
$27$ & $32{,}396$ & $25{,}756$ & $\sim 10^{15}$ \\
$36$ & $8{,}691$ & $6{,}860$ & $\sim 10^{21}$ \\
\midrule
\multicolumn{4}{l}{Industrial} \\
$\leq 20$ & $194{,}885$ & $239{,}674$ & $\le 2.4\times 10^{18}$ \\
\midrule
\multicolumn{4}{l}{Synthetic} \\
$10$ & $6{,}000$ & $3{,}000$ & $\approx 3.6\times 10^{6}$ \\
\bottomrule
\end{tabular*}
\end{table}

\begin{table*}[t]
\caption{Referee improvement over logged pages (\%), mean$_{\pm\mathrm{std}}$ over three seeds; Avg: unweighted mean across lengths.
Bold: best mean; $^{*}$: higher than every baseline on each seed.}\label{tab:main}
\small
\setlength{\tabcolsep}{2.7pt}
\begin{tabular}{lccccccccc}
\toprule
 & \multicolumn{4}{c}{KuaiRand-1K} & \multicolumn{4}{c}{RL4RS-SeqSlate} & Industrial \\
\cmidrule(lr){2-5}\cmidrule(lr){6-9}\cmidrule(lr){10-10}
Method & $L\le15$ & $L\le28$ & $L\le54$ & Avg & $L=18$ & $L=27$ & $L=36$ & Avg & $L\le20$ \\
\midrule
\multicolumn{10}{l}{\emph{Generative recommendation baselines}} \\
GoalRank~\cite{goalrank} & $23.67_{\pm0.33}$ & $25.48_{\pm0.40}$ & $25.02_{\pm1.35}$ & $24.73$ & $4.11_{\pm0.05}$ & $4.80_{\pm0.06}$ & $2.28_{\pm0.23}$ & $3.73$ & $12.30_{\pm0.23}$ \\
DeGRe~\cite{degre} & $24.66_{\pm0.42}$ & $27.04_{\pm0.16}$ & $28.38_{\pm0.57}$ & $26.69$ & $3.55_{\pm0.08}$ & $4.65_{\pm0.09}$ & $2.58_{\pm0.10}$ & $3.59$ & $12.75_{\pm0.13}$ \\
GFN4Rec~\cite{liu2023gfn4rec} & $23.49_{\pm0.20}$ & $28.86_{\pm0.32}$ & $31.40_{\pm0.58}$ & $27.92$ & $4.44_{\pm0.10}$ & $5.13_{\pm0.37}$ & $2.51_{\pm0.20}$ & $4.03$ & $11.01_{\pm0.24}$ \\
HAC~\cite{liu2023hac} & $13.89_{\pm0.99}$ & $15.11_{\pm0.44}$ & $18.46_{\pm0.79}$ & $15.82$ & $0.63_{\pm0.19}$ & $1.88_{\pm0.41}$ & $1.21_{\pm0.64}$ & $1.24$ & $1.61_{\pm0.89}$ \\
\midrule
\multicolumn{10}{l}{\emph{RL baselines}} \\
PPO~\cite{schulman2017proximal} & $26.42_{\pm0.11}$ & $32.98_{\pm0.36}$ & $38.85_{\pm0.47}$ & $32.75$ & $3.78_{\pm0.13}$ & $6.14_{\pm0.43}$ & $4.10_{\pm0.05}$ & $4.67$ & $10.44_{\pm0.29}$ \\
TD3N~\cite{fujimoto2018addressing} & $11.07_{\pm1.34}$ & $12.34_{\pm4.96}$ & $9.89_{\pm4.25}$ & $11.10$ & $0.87_{\pm0.63}$ & $2.21_{\pm1.31}$ & $1.03_{\pm0.40}$ & $1.37$ & $0.72_{\pm0.30}$ \\
GRPO~\cite{shao2024deepseekmath} & $25.91_{\pm0.22}$ & $33.03_{\pm0.45}$ & $39.28_{\pm0.42}$ & $32.74$ & $5.74_{\pm0.05}$ & $7.86_{\pm0.08}$ & $5.69_{\pm0.05}$ & $6.43$ & $12.68_{\pm0.65}$ \\
\midrule
Ours & $\mathbf{27.59}^{*}_{\pm0.30}$ & $\mathbf{34.49}^{*}_{\pm0.22}$ & $\mathbf{40.96}^{*}_{\pm0.28}$ & $\mathbf{34.35}$ & $\mathbf{6.82}^{*}_{\pm0.03}$ & $\mathbf{8.36}^{*}_{\pm0.04}$ & $\mathbf{6.09}^{*}_{\pm0.02}$ & $\mathbf{7.09}$ & $\mathbf{13.92}^{*}_{\pm0.01}$ \\
\bottomrule
\end{tabular}
\\[2pt]
{\footnotesize Referee score of the logged pages, per session: KuaiRand-1K $L\le15$: 8.23; KuaiRand-1K $L\le28$: 16.22; KuaiRand-1K $L\le54$: 31.31; RL4RS-SeqSlate $L=18$: 178.67; RL4RS-SeqSlate $L=27$: 331.28; RL4RS-SeqSlate $L=36$: 515.39. Industrial: absolute scores withheld.}
\end{table*}

\paragraph{Evaluators.}
On each public dataset, all methods train and select checkpoints using a fixed main evaluator trained only on the training split.
A separately trained referee, a learned model of user responses, scores the generated test pages.
To model responses in the test period, the referee is also trained on test-period logs, with validation sessions held out.
No method queries the referee or sees test logs during training or checkpoint selection, so there is no test-set leakage.
This follows simulator-based evaluation: HAC evaluates with a user response model pretrained on the full dataset~\cite{liu2023hac}, and KuaiSim uses response models pretrained on KuaiRand logs~\cite{zhao2023kuaisim}.
On held-out sessions, the main evaluator and referee reach click AUCs of 0.777 and 0.778 on KuaiRand-1K and purchase AUCs of 0.912 and 0.919 on RL4RS-SeqSlate (Appendix~\ref{app:evaluation-details}).
On the industrial dataset, the main evaluator is a listwise page-score model trained on the training day, and the referee for each test day is a version of the same model trained only on that day; neither is trained by us, and no method queries the referee. 

\paragraph{Baselines.}
We compare with generative recommendation methods GoalRank~\cite{goalrank}, DeGRe~\cite{degre}, GFN4Rec~\cite{liu2023gfn4rec} and HAC~\cite{liu2023hac}, and RL algorithms PPO~\cite{schulman2017proximal}, TD3N~\cite{fujimoto2018addressing} and GRPO~\cite{shao2024deepseekmath}.
TD3N denotes our adaptation of TD3 following the RL4RS benchmark code; continuous actor outputs are mapped to legal discrete candidates by a masked argmax.
Each baseline is adapted from its paper and official code, with implementation choices disclosed in Appendix~\ref{app:baseline-hp}.

\paragraph{Evaluation.}
All trained generators share the architecture and initial weights for a given seed, and results average three seeds unless reported separately.
Each run keeps the first checkpoint attaining its highest validation score, measured by greedy decoding with the evaluator used for training.
The trained generators are also decoded greedily on the same test requests.
On the public and industrial datasets, quality is improvement over logged pages:
$100\,(V_{\mathrm{method}}/V_{\mathrm{logged}}-1)$,
where both are mean test-page scores under the reported evaluator. \footnote{These are offline estimates on fixed candidate pools, not expected online benefits.}
Improvement is computed separately for each seed.
On synthetic data, quality is normalized ground-truth score,
$100\,N^{-1}\sum_{i=1}^{N}r(x_i)/r(x_i^*)$,
where $N$ is the number of test requests, $x_i$ the generated page and $x_i^*$ an optimal page for request $i$; 100 is optimal.

\paragraph{Our hyperparameters.}
The temperature is $\tau=1$ on both public datasets, $\tau=0.01$ on the industrial dataset, where page scores are normalized by the mean logged score, and $\tau=10^{-4}$ on synthetic data.
Each target batch is fitted with two epochs on the public and industrial datasets or two Adam steps on synthetic data.
Appendix~\ref{app:training-config} gives the full settings.

\subsection{Main results}\label{sec:main-results}

Whole-page policy improvement achieves the highest test-page quality at every length setting of both public datasets (Table~\ref{tab:main}).
Across lengths, its mean improvement over logged pages is 34.35\% on KuaiRand-1K and 7.09\% on RL4RS-SeqSlate, compared with the strongest baseline averages of 32.75\% and 6.43\%.

The larger-length setting of each dataset has the larger feasible-page space (Table~\ref{tab:dataset-space}).
On KuaiRand-1K, PPO and GRPO pull ahead of methods (e.g., DeGRe) that train toward fixed targets as capacity grows, and our lead persists against these stronger baselines at every capacity.
On RL4RS-SeqSlate, our lead also persists at the longest page length, where improvement over logged pages falls for all compared methods.

On the industrial dataset, our method improves served pages by 13.92\%, compared with 12.75\% for the strongest baseline (DeGRe), and leads every baseline on each seed.
The margin is about nine times DeGRe's seed standard deviation, and the lead holds on each of the three test days (14.03\%, 14.82\% and 12.91\%, versus 12.69\%, 13.55\% and 12.02\% for DeGRe).

\begin{figure*}[t]
\centering
\includegraphics[width=0.95\textwidth]{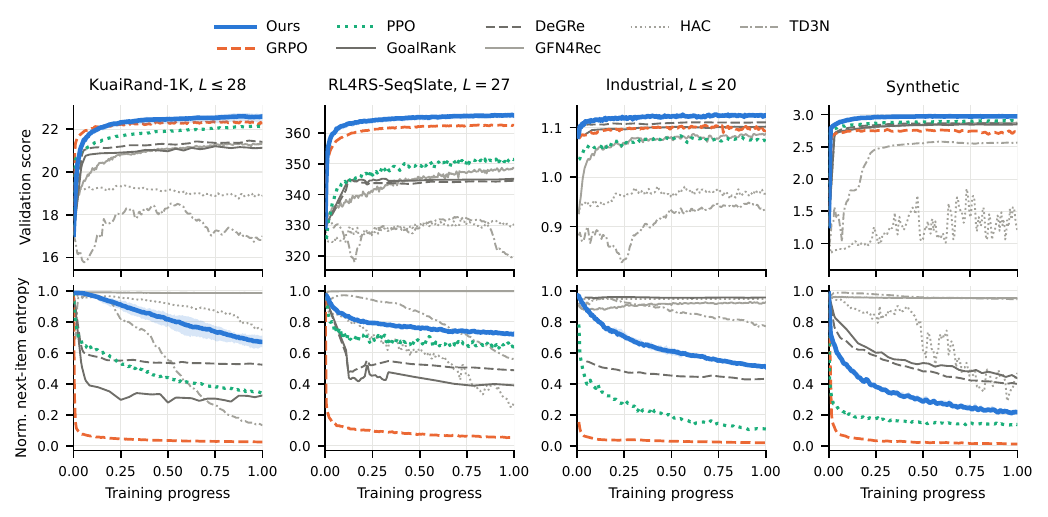}
\caption{Validation page score (top) and normalized next-item entropy (bottom) during training.
Progress is each run's step divided by its final step; lines are means over three seeds, and bands show our seed range.
Validation score is the main-evaluator score on the public and industrial datasets, normalized by the mean logged score on the industrial dataset, and the ground-truth score on synthetic data.
HAC/TD3N entropy measures action scores, not executed behavior.}
\label{fig:dynamics}
\end{figure*}

Figure~\ref{fig:dynamics} tracks validation score and next-item entropy, normalized by the log of the legal-set size, along greedy validation pages.
Our method reaches the highest validation score in every setting while retaining next-item entropy at the end of training: 0.67 and 0.72 on the public datasets and 0.51 on the industrial dataset.
This follows from the target construction: each round reweights the current next-item probabilities instead of replacing them, so candidates that look worse now keep some probability and can still be promoted in later rounds.
GRPO's next-item distributions become nearly deterministic early in training despite its entropy bonus (coefficient 0.01), ending at entropy 0.06 or below.

\subsection{Generation versus prefix-score beam search}\label{sec:serving-search}

\begin{table*}[!t]
\centering
\caption{Generator versus prefix-score beam search.
Score lift: referee improvement over logged pages (\%); evaluator calls and time are per page; time is the average batched generation time on one A100 80GB GPU, not online latency.
Generator: three-seed mean (it makes no evaluator calls at generation time); search: one deterministic run per width (1: greedy).
Bold: highest improvement.}\label{tab:evaluator-search}
\label{tab:evaluator-search-kr}\label{tab:evaluator-search-seq}
\small
\setlength{\tabcolsep}{4pt}
\begin{tabular}{lrrrrrrr}
\toprule
 & \multicolumn{3}{c}{KuaiRand-1K} & \multicolumn{3}{c}{RL4RS-SeqSlate} & Industrial \\
\cmidrule(lr){2-4}\cmidrule(lr){5-7}\cmidrule(lr){8-8}
Method & $L\le15$ & $L\le28$ & $L\le54$ & $L=18$ & $L=27$ & $L=36$ & $L\le20$ \\
\midrule
\multicolumn{8}{l}{\emph{Score lift (\%)}} \\
Our generator & $\mathbf{27.59}$ & $\mathbf{34.49}$ & $\mathbf{40.96}$ & $\mathbf{6.82}$ & $\mathbf{8.36}$ & $\mathbf{6.09}$ & $\mathbf{13.92}$ \\
Search width 1 & $24.20$ & $26.46$ & $28.45$ & $3.44$ & $4.99$ & $3.11$ & $9.94$ \\
Search width 8 & $26.37$ & $28.65$ & $30.49$ & $3.66$ & $5.14$ & $3.15$ & $12.85$ \\
Search width 32 & $27.38$ & $29.94$ & $31.73$ & $3.96$ & $5.31$ & $3.26$ & $13.65$ \\
\midrule
\multicolumn{8}{l}{\emph{Evaluator calls per page}} \\
Search width 1 & 137 & 518 & 2007 & 60 & 128 & 219 & 129 \\
Search width 8 & 974 & 3897 & 15564 & 432 & 957 & 1664 & 939 \\
Search width 32 & 3633 & 15431 & 62042 & 1616 & 3625 & 6404 & 3473 \\
\midrule
\multicolumn{8}{l}{\emph{Time (ms/page)}} \\
Our generator & 0.045 & 0.083 & 0.164 & 0.136 & 0.217 & 0.281 & 0.075 \\
Search width 1 & 0.801 & 1.750 & 6.736 & 2.183 & 3.894 & 6.752 & 1.543 \\
Search width 8 & 1.702 & 7.326 & 41.751 & 6.326 & 13.906 & 31.962 & 8.838 \\
Search width 32 & 4.822 & 26.626 & 162.505 & 19.804 & 47.756 & 116.550 & 30.801 \\
\bottomrule
\end{tabular}
\end{table*}

We compare greedy decoding from the trained generator with prefix-score beam search on the same test requests, candidate pools and page constraints (Table~\ref{tab:evaluator-search}).
Prefix-score beam search retains the best $W$ prefixes by their main-evaluator score, with $W\in\{1,8,32\}$; width 1 is greedy search.
Final pages are scored by the referee.
Generator results average three seeds; each search width is one deterministic run.
Computational costs are evaluator calls per page and average batched generation time per page.

Our greedily decoded generator achieves a higher mean page score under the referee than prefix-score beam search at every tested width and setting, without evaluator queries during generation; on the industrial dataset, it reaches 13.92\% versus 13.65\% for width 32.
Wider beams improve search quality with diminishing returns: quadrupling the width from 8 to 32 adds at most 1.29 points while roughly quadrupling evaluator calls, and our generator remains ahead even at width 32.
The measured average generation time is also lower for our generator at every setting.

Increasing beam width retains more alternatives, but each prefix is still ranked by the score of its placed items.
Our training instead compares next-item choices through complete pages sampled from the current generator.
The candidate-scoring ablation in \S\ref{sec:ablations} directly examines the benefit of including the remaining page in this supervision.
Scoring and timing protocols are detailed in Appendix~\ref{app:evaluator-search}; measured computational costs are compared within each dataset.

\subsection{Synthetic benchmark with known scores}\label{sec:exp-synth}

Scores on the public and industrial datasets are estimates from a learned referee, so they cannot show how close a generated page is to the best possible page.
On the synthetic benchmark, page scores are known exactly and optimal pages can be enumerated, which lets us measure quality against the optimum and separate where the score comes from.

\begin{table}[t]
\caption{Synthetic score and decomposition (\% of each request's optimal full-page score).
The individual-item contribution re-scores the same order without context-dependent click and browsing effects; page context is the remaining score.
Bold: largest Score and page-context contribution; $^{*}$: exceeds every baseline on each seed.}\label{tab:synth}
\small
\setlength{\tabcolsep}{3pt}
\begin{tabular}{lccc}
\toprule
Method & Score & \shortstack{Contribution\\from individual\\items} & \shortstack{Contribution\\from page\\context} \\
\midrule
\multicolumn{4}{l}{\emph{Generative recommendation baselines}} \\
GoalRank & $92.70_{\pm3.74}$ & $80.78$ & $11.92$ \\
DeGRe & $93.20_{\pm3.34}$ & $77.90$ & $15.29$ \\
GFN4Rec & $93.59_{\pm3.42}$ & $79.31$ & $14.28$ \\
HAC & $67.87_{\pm3.44}$ & $63.60$ & $4.27$ \\
\midrule
\multicolumn{4}{l}{\emph{RL baselines}} \\
PPO & $94.64_{\pm1.77}$ & $75.16$ & $19.48$ \\
TD3N & $81.54_{\pm9.75}$ & $76.36$ & $5.17$ \\
GRPO & $88.92_{\pm5.60}$ & $76.06$ & $12.85$ \\
\midrule
Ours & $\mathbf{97.16}^{*}_{\pm1.28}$ & $75.01$ & $\mathbf{22.15}^{*}$ \\
\bottomrule
\end{tabular}
\end{table}

\paragraph{Score decomposition.}
Our method reaches 97.16\% of the optimum, compared with PPO's 94.64\%, and leads every baseline on each seed.
Its higher page score comes from a larger page-context contribution, not a larger contribution from individual items (Table~\ref{tab:synth}).
To decompose these scores, we re-score the same generated orders with context-dependent click and browsing effects removed, retaining base click probabilities, item utilities and fixed position weights.
This item-separable score gives the contribution from individual items; subtracting it from the original page score gives the contribution from page context (Appendix~\ref{app:synthetic-details}).
Both contributions use the same full-simulator optimal-page denominator: ours has an individual-item contribution of 75.01\% versus PPO's 75.16\%, but a page-context contribution of 22.15\% versus 19.48\%.
An order that is optimal in the item-separable reference reaches only 75.9\% of the optimum in the full simulator, illustrating how the order preferred by individual-item scores can differ from the best whole-page order.

\subsection{Ablations}\label{sec:ablations}

We ablate two components on the public datasets: scoring candidates through complete pages, and constructing training data with the updated generator in each round.

\begin{figure}[t]
\centering
\includegraphics[width=0.95\columnwidth]{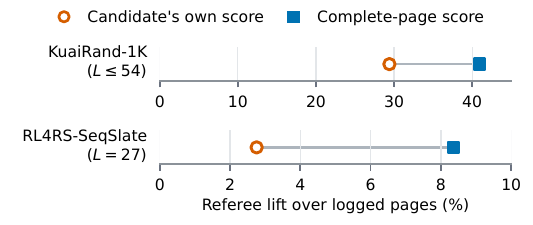}
\caption{Own score evaluates the candidate's slot given the prefix, without completion. Three-seed means of referee lift over logged pages (\%).}\label{fig:candidate-scoring}
\end{figure}

\paragraph{Scoring candidates through complete pages.}
We replace complete-page candidate scoring with the candidate's own slot score given the prefix.
On both public datasets, this yields clearly worse pages (Figure~\ref{fig:candidate-scoring}; Appendix~\ref{app:candidate-scoring}). A candidate's own score ignores how the candidate changes the contributions of the items placed after it, such as redundancy with or complementarity to later items.
The remaining page therefore matters for each next-item choice.

\begin{figure}[t]
\centering
\includegraphics[width=0.95\columnwidth]{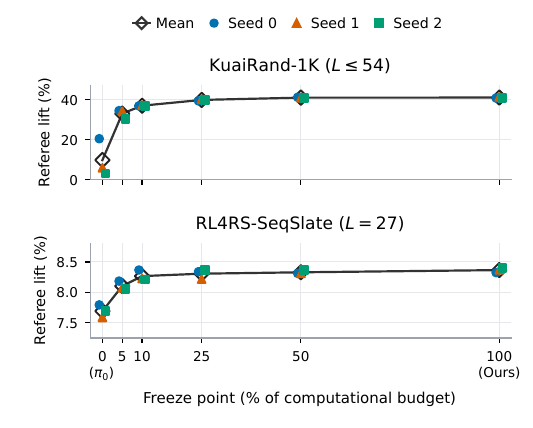}
\caption{Referee lift over logged pages (\%) versus freeze point. Colored points show individual seeds; diamonds show three-seed means. At 0\%, data construction uses $\pi_0$; 100\% is Ours.}\label{fig:freeze-point}
\end{figure}

\paragraph{Data construction with the updated generator.}
We ablate the use of the updated generator for data construction on RL4RS-SeqSlate at $L=27$ and KuaiRand-1K at $L\le54$ under matched computational budgets.
We freeze the data-construction generator after a chosen fraction of the computational budget, while the trained generator keeps fitting fresh targets to the full budget; 100\% denotes Ours (Figure~\ref{fig:freeze-point}).
Keeping the construction generator at $\pi_0$ lowers mean page quality, while the later freeze points yield means close to Ours.
The frozen generator supplies partial pages, candidate selection, completions and target-prior probabilities; Appendix~\ref{app:reference-comparison} reports the protocol.

\section{Conclusion}\label{sec:conclusion}

Whole-page policy improvement compares next-item alternatives through complete pages and refreshes these comparisons with the updated generator.
Exact per-position updates improve page score; repeated exact full updates converge.
On two public datasets, an industrial dataset of logged pages and a synthetic benchmark with known page scores, it outperforms generative recommendation and representative RL baselines, while its trained generator beats tested prefix-score beam search without serving-time evaluator queries.
Ablations on the public datasets support complete-page scoring and data construction with the updated generator.

\bibliographystyle{ACM-Reference-Format}
\bibliography{refs}

\appendix

\section{Proofs for Section~\ref{sec:method}}\label{app:proofs}

\subsection{Basic identities}\label{app:identities}

All statements hold per request; expectations over requests are dropped.
We use two standard tools.

We call partial pages states.
We write $r(x)$ for the complete-page evaluator score.
For a generator $q$, $V^{q}(s)$ is the expected evaluator score when $q$ completes the page from state $s$, so $V^{q}(s)=\sum_a q(a\mid s)Q^{q}(s,a)$, and $Q^{q}(s,a)$ is $V^{q}$ at the state $(s,a)$ that appends $a$ to $s$.

\begin{lemma}[Performance difference~\cite{kakade2002approximately}]\label{lem:pdl}
For any generators $p$ and $q$,
$J(p)-J(q)=\sum_{t=1}^{L}\E_{s\sim d_t^{p}}\big[\sum_a p(a\mid s)\,Q^{q}(s,a)-V^{q}(s)\big]$.
\end{lemma}

\begin{proof}
Sample a page from $p$ with states $s_1,\dots,s_{L+1}$, where $s_{L+1}=x$ and $V^{q}(x)=r(x)$.
Telescoping, $r(x)-V^{q}(s_1)=\sum_{t}\big[V^{q}(s_{t+1})-V^{q}(s_t)\big]$.
Since $s_{t+1}=(s_t,a_t)$ with $a_t\sim p(\cdot\mid s_t)$, $\E[V^{q}(s_{t+1})\mid s_t]=\sum_a p(a\mid s_t)Q^{q}(s_t,a)$.
Taking expectations under $p$ gives the claim.
\end{proof}

\begin{lemma}[Gibbs variational identity]\label{lem:gibbs}
Let $\mu$ be a distribution on a finite set, $f$ a function on it, $\tau>0$, $q\propto\mu\exp(f/\tau)$ and $Z=\sum_a\mu(a)\exp(f(a)/\tau)$.
For every $p$ with support in that of $\mu$,
$\sum_a p(a)f(a)-\tau\KL(p\|\mu)=\tau\log Z-\tau\KL(p\|q)$.
Hence $q$ is the unique maximizer of the left side, and
\begin{equation}\label{eq:gibbs2}
  \E_{q}f-\E_{\mu}f=\tau\big[\KL(q\|\mu)+\KL(\mu\|q)\big]\ge0,
\end{equation}
with equality if and only if $f$ is constant on the support of $\mu$.
\end{lemma}

\begin{proof}
$\log q=\log\mu+f/\tau-\log Z$, so $\tau\KL(p\|q)=\tau\KL(p\|\mu)-\E_p f+\tau\log Z$.
Evaluating the identity at $p=q$ and at $p=\mu$ and subtracting gives~\eqref{eq:gibbs2}.
Equality holds iff $q=\mu$, i.e.\ iff $f$ is constant on the support of $\mu$.
\end{proof}

\subsection{First-order agreement of the local objective with $J$}\label{app:surrogate}
At $\pi=\pi_k$, $\sum_a\pi_k(a\mid s)Q^{\pi_k}(s,a)-V^{\pi_k}(s)=0$ and the KL term is zero, so $\mathcal{L}_k(\pi_k)=J(\pi_k)$.
For a smooth, support-preserving parameterization $\pi_\theta$ with $\pi_{\theta_k}=\pi_k$, Lemma~\ref{lem:pdl} gives $J(\pi_\theta)-J(\pi_k)=\sum_t\E_{s\sim d_t^{\pi_\theta}}\big[\sum_a\pi_\theta(a\mid s)Q^{\pi_k}(s,a)-V^{\pi_k}(s)\big]$.
Differentiating at $\theta_k$, the term from $d_t^{\pi_\theta}$ multiplies $\sum_a\pi_k(a\mid s)Q^{\pi_k}(s,a)-V^{\pi_k}(s)=0$.
So $\nabla J(\theta_k)=\sum_t\E_{s\sim d_t^{\pi_k}}\big[\sum_a\nabla\pi_\theta(a\mid s)Q^{\pi_k}(s,a)\big]$, the gradient of the first term in the bracket of~\eqref{eq:surrogate}.
The KL term is minimized at $\pi_k$ and has zero gradient there.
Hence $\nabla\mathcal{L}_k=\nabla J$ at $\pi_k$.

\subsection{Proof of Proposition~\ref{prop:split}}\label{app:split}

\begin{proof}
Let $F_s(p)$ be the objective of~\eqref{eq:perstate}.
Then $\mathcal{L}_k(\pi)=J(\pi_k)+\sum_t\E_{s\sim d_t^{\pi_k}}\big[F_s(\pi(\cdot\mid s))-V^{\pi_k}(s)\big]$.
A state determines its position $t$, so each conditional $\pi(\cdot\mid s)$ enters one term, with weight $d_t^{\pi_k}(s)\ge0$.
Lemma~\ref{lem:gibbs} with $\mu=\pi_k(\cdot\mid s)$ and $f=Q^{\pi_k}(s,\cdot)$ gives $F_s(p)=\tau\log Z_k(s)-\tau\KL(p\|q_k(\cdot\mid s))$, uniquely maximized by $q_k(\cdot\mid s)$.
At states that $\pi_k$ does not reach we define $q_k$ by~\eqref{eq:target} as well.
Summing over states also gives the exact surrogate gap
\begin{equation}\label{eq:gap}
  \mathcal{L}_k(q_k)-\mathcal{L}_k(\pi)=\tau\sum_{t}\E_{s\sim d_t^{\pi_k}}\KL\big(\pi(\cdot\mid s)\,\|\,q_k(\cdot\mid s)\big),
\end{equation}
so maximizing $\mathcal{L}_k$ is minimizing a reverse KL to $q_k$ on the current generator's states.
Note also that $q_k(\cdot\mid s)$ and $\pi_k(\cdot\mid s)$ have the same support.
\end{proof}

\subsection{Proof of Proposition~\ref{prop:improve}}\label{app:improve}

\begin{proof}
Apply~\eqref{eq:gibbs2} at each state with $\mu=\pi_k(\cdot\mid s)$ and $f=Q^{\pi_k}(s,\cdot)$:
\[
  \begin{aligned}
  G_k(s)&:=\E_{a\sim q_k(\cdot\mid s)}Q^{\pi_k}(s,a)-V^{\pi_k}(s)\\
  &=\tau\big[\KL_s(q_k\|\pi_k)+\KL_s(\pi_k\|q_k)\big]\ge0.
  \end{aligned}
\]
Lemma~\ref{lem:pdl} with $p=q_k$ and $q=\pi_k$ gives $J(q_k)-J(\pi_k)=\sum_t\E_{s\sim d_t^{q_k}}G_k(s)$, which is~\eqref{eq:improve}.
\end{proof}

\subsection{Proof of Proposition~\ref{prop:renorm}}\label{app:renorm}
Let $q_k^S$ be~\eqref{eq:renorm} with $Q^{\pi_k}$, as in Section~\ref{sec:theory}.
Write $U(s)=\mathcal A(s)\setminus S(s)$ and $m(s)=\sum_{a\in S(s)}\pi_k(a\mid s)$.
For $m(s)>0$, let $\rho_s$ and $\tilde p_s$ be $\pi_k(\cdot\mid s)$ and $q_k^S(\cdot\mid s)$ restricted to $S(s)$ and renormalized.

\begin{proof}
Fix $s$ and write $S=S(s)$, $U=U(s)$, $m=m(s)$, $Q_a=Q^{\pi_k}(s,a)$.
If $m=0$, normalization and the fixed unscored probabilities force every feasible $p$ to equal $\pi_k(\cdot\mid s)$.
Consequently, $q_k^S(\cdot\mid s)=\pi_k(\cdot\mid s)$ and the one-position advantage is zero.
Assume $m>0$ below.
Every feasible $p$ puts mass $m$ on $S$, so we can write $p=m\,\tilde p$ and $\pi_k(\cdot\mid s)=m\,\rho$ on $S$, with $\tilde p$ and $\rho$ distributions on $S$.
On $U$, $p=\pi_k$, so $\sum_{a\in U}p(a)Q_a$ is a constant and the KL terms vanish.
On $S$, $\sum_{a\in S}p(a)\log\frac{p(a)}{\pi_k(a\mid s)}=m\KL(\tilde p\|\rho)$ and $\sum_{a\in S}p(a)Q_a=m\,\E_{\tilde p}Q$.
Hence $F_s(p)=\text{const}+m\big[\E_{\tilde p}Q-\tau\KL(\tilde p\|\rho)\big]$.
By Lemma~\ref{lem:gibbs} on $S$, this is maximized by $\tilde p\propto\rho\exp(Q/\tau)$, which is~\eqref{eq:renorm}.
For the improvement, the terms on $U$ cancel, and~\eqref{eq:gibbs2} on $S$ with $\mu=\rho$ gives
\[
  \begin{aligned}
  G_k^{S}(s)&:=\E_{q_k^{S}(\cdot\mid s)}Q-V^{\pi_k}(s)=m\big(\E_{\tilde p}Q-\E_\rho Q\big)\\
  &=m\,\tau\big[\KL(\tilde p\|\rho)+\KL(\rho\|\tilde p)\big]\ge0.
  \end{aligned}
\]
Lemma~\ref{lem:pdl} with $p=q_k^{S}$ and $q=\pi_k$ gives~\eqref{eq:renorm-improve}.
\end{proof}

\subsection{Convergence of exact target updates}\label{app:convergence}
We prove Proposition~\ref{prop:convergence} for the exact full-scoring update~\eqref{eq:target}.
The argument adapts the constant-step policy mirror descent analyses for discounted problems~\cite{xiao2022convergence,lan2022policy} to our finite-horizon page construction.
As elsewhere in this appendix, we fix a request and omit the expectation over requests.

\begin{proposition}[Exact-update rate]\label{prop:convergence-rate}
Let $R=\max_x r(x)-\min_x r(x)$ be the evaluator's score range over the request's feasible pages.
Suppose $\pi_{j+1}=q_j$ for all $j$, with a fixed $\tau>0$.
For any feasible page $x^*$ with $\pi_0(x^*)>0$ and any integer $K\ge1$,
\begin{equation}\label{eq:convergence-rate}
r(x^*)-J(\pi_K)
\le
\frac{\tau\log(1/\pi_0(x^*))+LR}{K}.
\end{equation}
\end{proposition}

\begin{proof}
First, the improvement proof applies from any partial page.
At every $s$, the one-position advantage
$\sum_a\pi_{j+1}(a\mid s)Q^{\pi_j}(s,a)-V^{\pi_j}(s)$
is nonnegative by~\eqref{eq:gibbs2}.
Lemma~\ref{lem:pdl}, started from $s$, therefore gives
$V^{\pi_{j+1}}(s)\ge V^{\pi_j}(s)$.
The same argument includes the root, so $J(\pi_j)$ is nondecreasing.
The target update also preserves support.

Write $x^*=(a_1^*,\ldots,a_L^*)$ and let $s_t^*$ be its partial page at position $t$.
Fix one such partial page $s=s_t^*$ and its next item $a^*=a_t^*$.
Define the normalizer
$Z_j(s)=\sum_a\pi_j(a\mid s)\exp(Q^{\pi_j}(s,a)/\tau)$.
Equation~\eqref{eq:target} gives
\begin{equation}\label{eq:convergence-local}
\begin{aligned}
Q^{\pi_j}(s,a^*)-V^{\pi_j}(s)
={}&\tau\log\frac{\pi_{j+1}(a^*\mid s)}{\pi_j(a^*\mid s)}+\tau\log Z_j(s)-V^{\pi_j}(s).
\end{aligned}
\end{equation}
All probabilities in this ratio are positive because $\pi_0(x^*)>0$ and support is preserved.

By Lemma~\ref{lem:gibbs},
\[
\begin{aligned}
\tau\log Z_j(s)
&=\sum_a\pi_{j+1}(a\mid s)Q^{\pi_j}(s,a)
-\tau\KL_s(\pi_{j+1}\|\pi_j)\\
&\le\sum_a\pi_{j+1}(a\mid s)Q^{\pi_j}(s,a)
\le V^{\pi_{j+1}}(s).
\end{aligned}
\]
The last inequality follows from
$Q^{\pi_j}(s,a)=V^{\pi_j}((s,a))\le V^{\pi_{j+1}}((s,a))$.
Substitute this inequality into~\eqref{eq:convergence-local} and sum over $j=0,\ldots,K-1$.
Both the log ratios and the $V^{\pi_j}$ differences telescope:
\[
\begin{aligned}
&\sum_{j<K}\big[Q^{\pi_j}(s,a^*)-V^{\pi_j}(s)\big]\\
&\quad\le
\tau\log\frac{\pi_K(a^*\mid s)}{\pi_0(a^*\mid s)}
+V^{\pi_K}(s)-V^{\pi_0}(s)\\
&\quad\le \tau\log(1/\pi_0(a^*\mid s))+R.
\end{aligned}
\]

Apply Lemma~\ref{lem:pdl} with the deterministic generator that follows $x^*$ and the generator $\pi_j$:
\[
r(x^*)-J(\pi_j)
=\sum_{t=1}^{L}
\big[Q^{\pi_j}(s_t^*,a_t^*)-V^{\pi_j}(s_t^*)\big].
\]
Summing over $j<K$ and over the $L$ partial pages gives
\[
\sum_{j<K}\big[r(x^*)-J(\pi_j)\big]
\le\tau\log(1/\pi_0(x^*))+LR,
\]
using $\pi_0(x^*)=\prod_t\pi_0(a_t^*\mid s_t^*)$.
Monotonicity of $J(\pi_j)$ bounds the left side below by
$K[r(x^*)-J(\pi_K)]$, proving~\eqref{eq:convergence-rate}.

Choose $x^*$ to maximize $r$ among pages supported by $\pi_0$.
Support preservation also gives $J(\pi_K)\le r(x^*)$.
The gap therefore tends to zero.
Full support over legal choices includes every feasible page and yields the global optimum.
\end{proof}

\section{Experimental details}\label{app:method-details}

\begin{table*}[t]
\caption{Reported baseline configurations.
$^{\dagger}$: selected on seed-0 validation.
``as KR'': the KuaiRand-1K configuration at $L\le28$.
Industrial: settings that differ from the shared ones; all industrial runs use batch 256.}\label{tab:baseline-hp}
\footnotesize
\setlength{\tabcolsep}{3pt}
\begin{tabular}{p{1.3cm}p{4.4cm}p{2.4cm}p{2.4cm}p{1.5cm}p{3.7cm}}
\toprule
Method & Shared settings & KuaiRand-1K & RL4RS-SeqSlate & Synthetic & Industrial \\
\midrule
GoalRank & page loss over 16 auxiliary lists and the current greedy list, weights $\mathrm{softmax}(z(r)/1)$; Adam $10^{-3}$, batch 256 & --- & --- & 32 epochs & 16 fixed evaluator-guided lists without the current greedy list; weights $\mathrm{softmax}(z(r)/0.3)$ \\
DeGRe & beam 8, all beams weighted by $\mathrm{softmax}(r/1)$, soft label 0.0099; Adam $10^{-3}$, batch 256 & --- & --- & 32 epochs & beam 32; hard label from the best beam \\
GFN4Rec & trajectory balance, list-average reward, $b_r=0.1$, $b_f=1$, $b_z=1$, batch 128, replay 50K & lr $10^{-4}$ & lr $3\cdot10^{-4\,\dagger}$ & lr $3\cdot10^{-4\,\dagger}$ & lr $3\cdot10^{-4}$; $b_r=0.5$, reward $(r+b_r)^{10}$; no replay \\
HAC & official three-optimizer structure; critic $10^{-3}$, behavior $10^{-4}$, alignment weight 0.1 (lr $10^{-4}$), Gaussian noise 0.1, $\epsilon=0.01$ & actor $10^{-5}$ / $3\cdot10^{-4}$ / $10^{-5\,\dagger}$ & actor $10^{-5\,\dagger}$ & actor $3\cdot10^{-4\,\dagger}$ & actor $3\cdot10^{-4}$, with behavior and alignment losses in the actor optimizer; alignment weight 1.0; noise 0.3; no $\epsilon$ \\
PPO & entropy 0.01 & clip 0.2, GAE 0.95, lr $10^{-4}$ / $10^{-4}$ / $3\cdot10^{-4}$ & RL4RS PPO: clip 0.3, GAE 1.0, adaptive KL, lr $10^{-4}$ & as KR & as KR, lr $3\cdot10^{-4}$, entropy 0.001 \\
TD3N & twin critics, smoothing 0.2 (clip 0.5), delay 2, critic $10^{-3}$, Gaussian noise 0.1, 10K random warm-up steps & actor $10^{-3}$ & actor $10^{-3\,\dagger}$ & actor $10^{-4\,\dagger}$ & actor $10^{-4}$, critic $3\cdot10^{-4}$; noise 0.3; no warm-up \\
GRPO & 8 pages per request, group-normalized advantage, clip 0.2, 2 epochs, entropy 0.01 & lr $3\cdot10^{-4}$ & lr $5\cdot10^{-4}$ / $3\cdot10^{-4}$ / $3\cdot10^{-4}$ & lr $3\cdot10^{-4}$ & lr $10^{-4}$; dual-clipped ratio \\
\bottomrule
\end{tabular}
\end{table*}
\subsection{Training configuration}\label{app:training-config}\label{app:sensitivity}\label{app:target-implementation}
All methods share a Transformer pointer generator (hidden size 64, four heads, two encoder and two decoder layers, about 250K parameters), randomly initialized for each seed without pretraining.
Table~\ref{tab:method-settings} lists our settings; $\tau$ is selected on validation data.
Each round draws 6{,}000 training requests on the public datasets (30{,}000 on the industrial dataset) and fits two epochs at batch size 32 (256); the synthetic benchmark takes $\UpdatesSynth$ Adam steps per 32-page target batch.

\subsection{Evaluators}\label{app:evaluation-details}
Both evaluators are Transformer page scorers: on KuaiRand-1K, causal with a history encoder, predicting click, long view and like with page weights 1, 1 and 10; on RL4RS-SeqSlate, causal across slates and bidirectional within a slate, predicting purchase probability times an item-level weight.

\begin{table}[t]
\caption{Per-dataset settings for whole-page policy improvement; legal items are counted per state.}\label{tab:method-settings}
\small
\setlength{\tabcolsep}{3pt}
\begin{tabular}{lcccc}
\toprule
Dataset & $\kappa$, $\kappa_{\mathrm r}$ & $\tau$ & Legal items & Fit \\
\midrule
KuaiRand-1K & 6, 2 & 1 & 15 / 29 / 56 & \EpochsKR{} epochs \\
RL4RS-SeqSlate & 6, 2 & 1 & $\le$6 / 9 / 12 & \EpochsSeq{} epochs \\
Industrial & 16, 2 & 0.01 & $\le$20 & 2 epochs \\
Synthetic & 6, 2 & $10^{-4}$ & $11-t$ & $U=\UpdatesSynth$ steps \\
\bottomrule
\end{tabular}
\end{table}

\subsection{Data partitions and constraints}\label{app:dataset-constraints}
\paragraph{KuaiRand-1K}
A page joins $k\in\{2,4,8\}$ consecutive requests within a session (30-minute gaps), excluding pages above the 99.5th length percentile or with fewer than two items.
Training, validation and test pages start through May 1, on May 2--4 and on May 5--8, 2022.
Extra candidates are up to $n$ items the user saw within the next 24 hours, excluding items already on the page or in the evaluator's history, sampled with a page-ID seed so that all methods and seeds share the pools.

\paragraph{RL4RS-SeqSlate.}
Session ID modulo 10 assigns dataset-B sessions to training (0--4), validation (5) and test (6--9); each $L$ keeps sessions with exactly $L/9$ slates and a legal logged order.
Following the official code, each entry keeps its row eligibility, item IDs cannot repeat within a slate, and each slate holds at most one special item.

\subsection{Synthetic simulator}\label{app:synthetic-details}
Each seed defines a world with 5{,}000 items, 30{,}000 requests, four item formats and ten topics, split 80/10/10; runs use 6{,}000 training, 500 validation and all 3{,}000 test requests.
Items have a topic vector $\theta_i$ (one or two topics), an embedding $E_i$ near their topic centers, a bias $b_i$, a utility parameter $w_i$ and a satisfaction parameter $h_i$.
A request has an embedding $U$ and format preferences $U_f$, and keeps the top ten of 40 sampled items by noisy relevance.
For a page $(i_1,\ldots,i_{10})$, with position bucket $b(t)=\lfloor4(t-1)/10\rfloor$ and $\delta(d)=e^{-(d-1)/2}$, the click logit is
\[
\begin{aligned}
z_t={}&U^\top E_{i_t}+U_f[f_{i_t}]+b_{i_t}-0.5w_{i_t}+A[f_{i_t},b(t)]\\
&-2\textstyle\sum_{k}\theta_{i_tk}\log\!\big(1+\sum_{j<t}\theta_{i_jk}\big)+2\textstyle\sum_{j<t}\delta(t-j)\theta_{i_j}^\top M\theta_{i_t}-\mathrm{run}_t,
\end{aligned}
\]
where $A$ is a random format-by-position matrix, $M$ pairs complementary topics and $\mathrm{run}_t$ counts preceding consecutive items of the same format.
The click probability at a reached position is $c_t=t^{-1/2}\sigma(z_t)$, a clicked item has utility $v_i=\exp(w_i-1/2)$, and $r(x)=\sum_t e_tc_tv_{i_t}$, where $e_t$ is the probability of reaching position $t$.
Users stop after a click with probability $(1-e^{-0.5})\sigma(h_i)$, after a non-click with probability $0.05(1-e^{-0.5})$, and abandon after a three-slot screen without clicks with probability $1-e^{-2}$; $e_t$ is computed exactly by a two-state recursion.
The exact optimum enumerates all $10!$ orders.
The item-separable reference removes the four page-context terms and replaces $e_t$ by its average $\bar e_t$ over logged pages: $r_{\mathrm{sep}}(x)=\sum_t\bar e_tt^{-1/2}\sigma(z_0(i_t))v_{i_t}$ with the base logit $z_0$.
The page-context contribution is $r(x)-r_{\mathrm{sep}}(x)$; both parts are normalized by the optimal-page score.

\section{Baseline implementations}\label{app:baseline-hp}\label{app:baseline-protocol}\label{app:baseline-selection}
Baselines use the shared generator, masks, initialization and evaluator, which replaces each method's own reward model; Table~\ref{tab:baseline-hp} lists their settings.
Table~\ref{tab:training-evaluator-calls} reports training-time evaluator calls; a call scores one complete or partial page, and GoalRank and DeGRe include their offline target construction.

\begin{table}[t]
\centering
\caption{Training-time evaluator calls (millions).}\label{tab:training-evaluator-calls}
\footnotesize
\setlength{\tabcolsep}{1.5pt}
\begin{tabular}{lrrrrrrr}
\toprule
Method & \multicolumn{3}{c}{KuaiRand-1K} & \multicolumn{3}{c}{RL4RS-SeqSlate} & Industrial \\
\cmidrule(lr){2-4}\cmidrule(lr){5-7}\cmidrule(lr){8-8}
 & $L\le15$ & $L\le28$ & $L\le54$ & $L=18$ & $L=27$ & $L=36$ & $L\le20$ \\
\midrule
GoalRank & 432.9 & 741.5 & 1242.5 & 89.3 & 21.3 & 9.5 & 127.5 \\
DeGRe & 1302.5 & 2333.4 & 3960.8 & 256.7 & 64.0 & 29.4 & 697.8 \\
GFN4Rec & 3.2 & 3.4 & 2.6 & 3.3 & 3.5 & 2.1 & 3.8 \\
HAC & 2.7 & 2.7 & 2.7 & 3.1 & 3.6 & 2.7 & 3.1 \\
PPO & 11.3 & 17.7 & 13.1 & 8.9 & 6.1 & 4.0 & 3.1 \\
TD3N & 3.7 & 2.8 & 3.6 & 2.7 & 3.4 & 2.7 & 4.0 \\
GRPO & 13.4 & 21.3 & 20.7 & 10.9 & 28.0 & 17.7 & 14.2 \\
\midrule
Ours & 66.6 & 168.9 & 397.1 & 62.1 & 155.3 & 159.9 & 396.1 \\
\bottomrule
\end{tabular}
\end{table}

\subsection{Implementation choices}\label{app:baseline-disclosures}
\begin{enumerate}[leftmargin=*]
\item GoalRank's official recipe uses learning rate $10^{-4}$, batch 512 and 200 epochs.
\item HAC uses alignment learning rate $10^{-4}$ instead of the unstable official 0.1; HAC and TD3N use Gaussian noise instead of the official codes' uniform and Ornstein--Uhlenbeck noise.
\item On the industrial dataset, GRPO uses a dual-clipped ratio and a learning rate of $10^{-4}$ to avoid training divergence.
\end{enumerate}

\section{Ablation and search protocols}\label{app:results}\label{app:real-results}
\subsection{Candidate-scoring protocol}\label{app:candidate-scoring}
The own-score arm keeps our hyperparameters, scored-set rule, target formula, fitting recipe, seeds, computational budgets and request schedule and removes the completion rollouts.
On RL4RS-SeqSlate, the own score is the candidate slot's predicted score with a padded tail.

\subsection{Data construction with updated generator}\label{app:reference-comparison}
Figure~\ref{fig:freeze-point} reports the comparisons.
The ablated variant uses $\pi_0$, or a frozen $\pi_m$, wherever data construction draws on the generator: sampled partial pages, scored-candidate selection, completions and prior probabilities in the target weights.
It still constructs fresh targets throughout training, and the trained generator is never reset.
We reuse the reported runs and fix each ablated run's computational budget to the matching reported run's number of rounds, with the same per-round request sets, Adam steps per round and checkpoint-selection rule; evaluator-call totals match, within 0.2\% on RL4RS-SeqSlate.

For a freeze point $f$ and computational budget $T$, the data-construction generator is fixed to $\pi_m$, $m=\operatorname{round}(fT/100)$, from the reference trajectory, and the trained generator continues from that trajectory's state for the remaining $T-m$ rounds; 100\% is the reported method.
A few ablated runs select checkpoints near the end of the budget, so these are comparisons at a matched computational budget, not asymptotic limits.

\subsection{Prefix-score beam search at serving time}\label{app:evaluator-search}
Search uses the candidate pools, legality masks and test requests of Table~\ref{tab:main}; widths $\{1,8,32\}$ were fixed in advance, and each width runs once per setting.
At each position, it extends every retained prefix with each legal next item and keeps the top $W$ by the main evaluator's score of the placed items; unfilled slots are masked, and no estimate of the remaining score is used.
On KuaiRand-1K the evaluator is causal, so this score is exactly the sum of the placed slots' contributions; on RL4RS-SeqSlate it is causal across slates but bidirectional within a slate.
Final pages are scored by the referee; infeasible search outputs, at most 0.63\% on RL4RS-SeqSlate, are replaced by logged pages.
Times are average batched generation times per page on one A100 80GB GPU, with CUDA synchronization around generation; they include candidate generation and feature construction and exclude data and model loading.
These are batch-amortized per-page times rather than single-request latencies, and are compared only within each dataset.

\end{document}